\documentclass[11pt,a4paper]{article}
\usepackage{authblk}
\usepackage{lmodern}
\usepackage{jstyle}
\usepackage{amsmath}
\usepackage{mathtools}
\usepackage{tikz}
\usepackage{empheq}
\usepackage{enumitem}
\usepackage{graphics}
\usepackage{wrapfig}
\usepackage{caption}
\usepackage{natbib}
\usepackage{xcolor}
\usepackage{framed}
\usepackage{array}
\definecolor{shadecolor}{gray}{0.925}

\usepackage[cbgreek]{textgreek}

\def\sideremark#1{\ifvmode\leavevmode\fi\vadjust{\vbox to0pt{\vss
 \hbox to 0pt{\hskip\hsize\hskip1em
 \vbox{\hsize3cm\tiny\raggedright\pretolerance10000
 \noindent #1\hfill}\hss}\vbox to8pt{\vfil}\vss}}}%

\newcommand{\bi}{\begin{itemize}}
\newcommand{\ei}{\end{itemize}}
\newcommand{\bea}{\begin{align}}
\newcommand{\eea}{\end{align}}
\newcommand{\be}{\begin{equation}}
\newcommand{\ee}{\end{equation}}

\newcommand{\tcb}{\textcolor{blue}}

\makeatletter
\renewcommand*\env@matrix[1][\arraystretch]{%
  \edef\arraystretch{#1}%
  \hskip -\arraycolsep
  \let\@ifnextchar\new@ifnextchar
  \array{*\c@MaxMatrixCols c}}
\makeatother

\author[\ensuremath{\nu}]{Charlotte SLEIGHT\footnote{Also at the Universit\'e Libre de Bruxelles and International Solvay Institutes, Belgium.}}
\affiliation[\ensuremath{\nu}]{School of Natural Sciences, Institute for Advanced Study,\\
1 Einstein Drive, Princeton, NJ 08540}

\author[\ensuremath{\mathsf{s}},\ensuremath{\mathsf{t}},\ensuremath{\mathsf{u}}]{\quad Massimo TARONNA}

\affiliation[\ensuremath{\mathsf{s}}]{Dipartimento di Fisica ``Ettore Pancini'', Universit\`a degli Studi di Napoli Federico II, \\Monte S. Angelo, Via Cintia, 80126 Napoli, Italy}

\affiliation[\ensuremath{\mathsf{t}}]{INFN, Sezione di Napoli, Monte S. Angelo, Via Cintia, 80126 Napoli, Italy}

\affiliation[\ensuremath{\mathsf{u}}]{Department of Physics, Princeton University,\\
Jadwin Hall, Princeton, NJ 08544}

\emailAdd{csleight@ias.edu, massimo.taronna@unina.it}

\title{\centering \LARGE From AdS to dS Exchanges: Spectral Representation, Mellin Amplitudes and Crossing}

\abstract{We present a simple general relation between tree-level exchanges in AdS and dS. This relation allows to directly import techniques and results for AdS Witten diagrams, both in position and momentum space, to boundary correlation functions in dS. In this work we apply this relation to define Mellin amplitudes and a spectral representation for exchanges in dS. We also derive the conformal block decomposition of a dS exchange, both in the direct and crossed channels, from their AdS counterparts. The relation between AdS and dS exchanges itself is derived using a recently introduced Mellin-Barnes representation for boundary correlators in momentum space, where (A)dS exchanges are straightforwardly fixed by a combination of factorisation, conformal symmetry and boundary conditions.}

\begin{document}

\begin{flushright}    
\texttt{}
\end{flushright}

\maketitle

\section{Introduction}\label{sec::Intro}

The last decade has seen significant progress in our understanding of scattering in anti-de Sitter (AdS) space. Through the AdS/CFT correspondence \cite{Maldacena:1997re} we can reformulate scattering processes in AdS in terms of correlation functions in Conformal Field Theory (CFT), which are sharply defined by the requirements of Conformal Symmetry, Unitarity and a consistent Operator Product (OPE) expansion. Accordingly, numerous highly effective techniques for the study of scattering in AdS have been developed within the Conformal Bootstrap programme, including: Mellin amplitudes \cite{Mack:2009mi,Penedones:2010ue}; Spectral Representation \cite{Dobrev:1977qv,Costa:2014kfa}; Large Spin Expansion \cite{Komargodski:2012ek,Fitzpatrick:2012yx,Alday:2016njk}; dispersion relations and inversion formulae \cite{Caron-Huot:2017vep,Simmons-Duffin:2017nub,Carmi:2019cub}.

With this note we take some simple steps towards extending some of these techniques to scattering in de Sitter (dS) space. Conformal Symmetry has indeed been effective in constraining boundary observables in dS in numerous previous works \cite{Antoniadis:2011ib,Creminelli:2011mw,Bzowski:2011ab,Mata:2012bx,Farrow:2018yni}, most recently within the Cosmological Bootstrap programme \cite{Arkani-Hamed:2015bza,Arkani-Hamed:2017fdk,Arkani-Hamed:2018kmz,Sleight:2019mgd,Sleight:2019hfp,Baumann:2019oyu,Baumann:2020dch,Green:2020ebl} which at the same time aims to import techniques and ideas from the scattering amplitudes in flat space. Our understanding of the properties required of consistent correlation functions in Euclidean CFTs dual to physics in dS is, however, very primordial compared to that which we enjoy in AdS. In this work we aim to take some small (though, we hope, useful) steps towards closing the gap between the Conformal Bootstrap and Cosmological Bootstrap programmes. In particular, using the Mellin-Barnes representation for momentum space boundary correlators introduced in \cite{Sleight:2019mgd,Sleight:2019hfp}, we strengthen the connection between in-in correlators on the late-time boundary of dS and Witten diagrams in AdS. This allows us to present a simple general relation \eqref{dSexch} between exchanges in dS and AdS at tree-level, allowing us to directly import AdS techniques (both in position- and momentum-space\footnote{See e.g. \cite{Raju:2010by,Raju:2012zr,Isono:2018rrb,Isono:2019wex,Albayrak:2019asr,Albayrak:2019yve,Albayrak:2020isk} for studies of Witten diagrams in momentum-space, as well as e.g. \cite{Maldacena:2011nz,Anninos:2014lwa,Ghosh:2014kba,Goon:2018fyu,Hillman:2019wgh} in dS.}) to in-in correlators in dS, which we briefly explore in the final section of this work. This includes: Spectral representation \S \ref{subsec::spectralrep} and Mellin amplitudes \S \ref{subsec::Mellinamplitudes} for dS exchanges, as well as their Conformal Block decomposition both in the direct and crossed channels \S \ref{subsec::crossing}.

The Mellin-Barnes representation \cite{Sleight:2019mgd,Sleight:2019hfp} itself provides us a new systematic understanding of boundary correlators in AdS and dS. As we shall see, it allows to straightforwardly reconstruct an exchange diagram from its on-shell part. The latter is fixed by a combination of factorisation, conformal symmetry and boundary conditions which, as we shall see, can be seamlessly implemented using the Mellin-Barnes representation. 

In this note we aim to avoid technical clutter, which for the intrepid reader we shall present in full glory elsewhere \cite{ToAppear2}.

\paragraph{Notation and Conventions} Whenever referring to the bulk of EAdS and dS we use Poincar\'e coordinates:
\begin{align}\label{poincare}
    ds^2_{\text{EAdS}}=\frac{dz^2+d{\bf x}^2}{z^2}\,, \qquad ds^2_{\text{dS}}=\frac{-d\eta^2+d{\bf x}^2}{\eta^2}\,,
\end{align}
where $z\in[0,\infty)$ and $\eta\in(-\infty,0]$. The three-dimensional vector ${\bf x}$ parameterises the boundary directions and ${\bf k}$ will denote the boundary momentum. Overall momentum conserving delta functions will always be left implicit. Often we will parameterise scaling dimensions $\Delta_\pm$ and $\Delta_j$ as: $\Delta_\pm = \frac{d}{2}\pm i \mu$ and $\Delta_j = \frac{d}{2}+i \mu_j$. Unless specified the latter scaling dimensions will be kept generic throughout, though ultimately one should restrict to the unitary values appropriate to the space-time (AdS or dS) under consideration.
 
\section{(A)dS Propagators}
\label{sec::prop}

In this section we review some relevant aspects of propagators in (A)dS. 

\paragraph{Bulk-Boundary Propagators.} The in-in bulk-boundary propagators in dS for the $\Delta=\Delta_\pm$ modes in Poincar\'e coordinates \eqref{poincare} can be obtained from their counterparts in EAdS through the analytic continuation $z=-\eta\, e^{\pm \frac{i\pi}{2}}$:
\begin{equation}\label{analbubo}
     K^{\text{dS}, \pm}_{\Delta,J}\left(\eta, {\bf k}\right)= K^{\text{AdS}}_{\Delta,J}\left(-\eta\, e^{\pm \frac{i\pi}{2}}, {\bf k}\right).
\end{equation}
Bulk-boundary propagators in EAdS for scalar fields are given by a modified Bessel function of the second kind in Poincar\'e coordinates \cite{Gubser:1998bc}. Bulk-boundary propagators for spinning fields (see e.g. \cite{Raju:2010by}) with the same $\Delta$ can be obtained from the latter by acting with a differential operator in the boundary momentum ${\bf k}$ \cite{Sleight:2016hyl}.

Using the Mellin-Barnes representation of bulk-boundary propagators, the analytic continuation \eqref{analbubo} reduces to multiplying the AdS bulk-boundary propagator by a simple phase. For scalar fields the Mellin-Barnes representation reads \cite{Sleight:2019mgd,Sleight:2019hfp}:
\begin{subequations}\label{MBbubosc}
\begin{align}
\hspace*{-0.6cm}K^{\text{AdS}}_{\Delta,J}\left(z, {\bf k}\right)&=\int^{+i\infty}_{-i\infty}\left[ds\right]\,K^{\text{AdS}}_{\Delta,J}\left(z, {\bf k}, s\right),\\
 \hspace*{-0.6cm}   K^{\text{AdS}}_{\Delta,0}\left(z, {\bf k}, s\right)&=\frac{z^{d-\Delta}}{2\sqrt{\pi}\,\Gamma\left(\Delta-\frac{d}{2}\right)}\,\Gamma\left(s+\tfrac{1}2\left(\Delta-\tfrac{d}{2}\right)\right)\Gamma\left(s-\tfrac{1}2\left(\Delta-\tfrac{d}{2}\right)\right)\left(\tfrac{z |\bold{k}|}2\right)^{2s+\Delta-\frac{d}{2}},
\end{align}
\end{subequations}
where $\left[ds\right]=\frac{ds}{2\pi i}$. We often refer to variables $s$ as an external Mellin variables, which are associated to external legs. From this one can write down the relation, 
\begin{equation}\label{dSAdSbubo}
   K^{\text{dS}, \pm}_{\Delta,J}\left(\eta, {\bf k}, s\right)=e^{\mp \left(s+\tfrac{1}2\left(\Delta-\tfrac{d}{2}\right)\right)\pi i}K^{\text{AdS}}_{\Delta,J}\left(-\eta, {\bf k},s\right).
\end{equation}
This also holds for fields of spin-$J$, which can be understood from the relation between spinning and scalar bulk-boundary propagators given in \cite{Sleight:2016hyl}.

\paragraph{Bulk-Bulk Propagators.} For bulk-bulk propagators it is useful to consider a special class of bi-local Eigenfunctions of the Laplacian,
\begin{equation}\label{harmadseom}
    \left(\nabla^2_1-m^2\right) \Omega^{\text{AdS}}_{\mu,J}\left(x_1;x_2\right)=0, \qquad m^2=-\left(\Delta_+\Delta_-+J\right), \qquad \Delta_\pm = \frac{d}{2}\pm i\mu,
\end{equation}
which are non-singular in the short distance limit. In momentum space on the boundary these are given by the following product of bulk-boundary propagators \cite{Leonhardt:2003qu,Costa:2014kfa}: 
\begin{equation}\label{AdSharm}
    \Omega^{\text{AdS}}_{\mu,J}\left(z_1,{\bf k};z_2,-{\bf k}\right) = \frac{\mu^2}{\pi} K^{\text{AdS}}_{\Delta_+,J}\left(z_1, {\bf k}\right)K^{\text{AdS}}_{\Delta_-,J}\left(z_2, -{\bf k}\right),
\end{equation}
often referred to as the split representation in the literature. Such harmonic functions are a complete basis for bi-local functions in AdS \cite{Costa:2014kfa}. The bulk-bulk propagator for the normalisable i.e. $\Delta_+$ boundary condition admits the following spectral representation \cite{Cornalba:2007fs,Costa:2014kfa}:
\begin{equation}\label{specpropAdSdirich}
    \Pi^{\text{AdS}}_{\Delta_+,J}\left(z_1,{\bf k};z_2,-{\bf k}\right) = \int^{\infty}_{-\infty} \frac{d\nu}{\nu^2+\left(\Delta_+-\tfrac{d}{2}\right)^2}\, \Omega^{\text{AdS}}_{\nu,J}\left(z_1,{\bf k};z_2,-{\bf k}\right)+\dots\,,
\end{equation}
where the $\ldots$ denote lower spin contact terms. For the non-normalisable boundary condition $\Delta_-$ the spectral integral is supplemented by a single harmonic function \eqref{AdSharm},\footnote{This is because harmonic functions on the Principal series $\mu =\nu \in \mathbb{R}$ are only a complete basis for normalisable functions.}
\begin{multline}\label{specpropAdSdeu}
    \Pi^{\text{AdS}}_{\Delta_-,J}\left(z_1,{\bf k};z_2,-{\bf k}\right) = \frac{2\pi i}{\mu}\Omega^{\text{AdS}}_{\mu,J}\left(z_1,{\bf k};z_2,-{\bf k}\right) \\ +\int^{\infty}_{-\infty} \frac{d\nu}{\nu^2+\left(\Delta_+-\tfrac{d}{2}\right)^2}\, \Omega^{\text{AdS}}_{\nu,J}\left(z_1,{\bf k};z_2,-{\bf k}\right)+\dots\,.
\end{multline}
Propagators for the $\Delta_\pm$ modes are placed on an equal footing by adopting the Mellin-Barnes representation,
\begin{equation}
   \Pi^{\text{AdS}}_{\Delta_\pm,J}\left(z_1,{\bf k};z_2,-{\bf k}\right) = \int^{i\infty}_{-i\infty}\frac{du d{\bar u}}{\left(2\pi i\right)^2}\,  \Pi^{\text{AdS}}_{\Delta_\pm,J}\left(z_1,{\bf k},u;z_2,-{\bf k},{\bar u}\right),
\end{equation}
where the spectral integral can be lifted \cite{Sleight:2019hfp}:
\begin{multline}\label{propAdSsum}
    \Pi^{\text{AdS}}_{\Delta_\pm,J}\left(z_1,{\bf k},u;z_2,-{\bf k},{\bar u}\right) = \csc\left(\pi\left(u+{\bar u}\right)\right) \omega_{\Delta_\pm}\left(u,{\bar u}\right)\\\times\Gamma(i\mu)\Gamma(-i\mu)\Omega^{\text{AdS}}_{\mu,J}\left(z_1,{\bf k},u;z_2,-{\bf k},{\bar u}\right),
\end{multline}
with
\begin{equation}\label{w+-}
    \omega_{\Delta_\pm}\left(u,{\bar u}\right) = 2 \sin\left(u\mp\tfrac{i\mu}{2}\right)\sin\left({\bar u}\mp\tfrac{i\mu}{2}\right).
\end{equation}
We often refer to $u$ and ${\bar u}$ as internal Mellin variables, which are associated to an internal leg. The Mellin-Barnes representation of the harmonic function \eqref{AdSharm} is inherited from that of its constituent bulk-boundary propagators,
\begin{equation}\label{mbharm}
    \Omega^{\text{AdS}}_{\mu,J}\left(z_1,{\bf k},u;z_2,-{\bf k},{\bar u}\right)=\frac{\mu^2}{\pi} K^{\text{AdS}}_{\Delta_+,J}\left(z_1, {\bf k},u\right)K^{\text{AdS}}_{\Delta_-,J}\left(z_2, -{\bf k},{\bar u} \right).
\end{equation}

The discontinuity of \eqref{propAdSsum} with respect to ${\sf s}=|{\bf k}|^2$ gives the corresponding on-shell propagator, which for the $\pm$ modes reads 
\begin{multline}
     \text{Disc}_{{\sf s}}\left[\Pi^{\text{AdS}}_{\Delta_\pm,J}\left(z_1,{\bf k},u;z_2,-{\bf k},{\bar u}\right)\right] \\=  \omega_{\Delta_\pm}\left(u,{\bar u}\right)\Gamma(i\mu)\Gamma(-i\mu)\Omega^{\text{AdS}}_{\mu,J}\left(z_1,{\bf k},u;z_2,-{\bf k},{\bar u}\right),
\end{multline}
and encodes the physical exchanged single-particle state. The contact terms in the exchange are instead encoded by the factor $\csc\left(\pi \left(u+{\bar u}\right)\right)$ in the full propagator, which is killed by the discontinuity through the identity
\begin{equation}
      \text{Disc}_{{\sf s}}\left[|{\bf k}|^{2\left(u+{\bar u}\right)}\right]=\sin \left(\pi\left(u+{\bar u}\right)\right)|{\bf k}|^{2\left(u+{\bar u}\right)}.
\end{equation}
In other words, at the level of the Mellin-Barnes representation the full propagator is reconstructed from its on-shell part by multiplying with the factor $\csc\left(\pi \left(u+{\bar u}\right)\right)$.

The functions $\omega_{\Delta_\pm}\left(u,{\bar u}\right)$ can be regarded as projectors onto the $\Delta_\pm$ modes. A single harmonic function \eqref{AdSharm}, being symmetric under $\mu \to -\mu$, is a specific linear combination of $\Delta_\pm$ modes. The zeros of $\omega_{\Delta_\pm}\left(u,{\bar u}\right)$ overlap with the poles in the Mellin-Barnes representation \eqref{mbharm} of the harmonic function that generate the $\Delta_\mp$ mode contributions, thus projecting them out. The Mellin-Barnes representation of the most general propagator in AdS is a linear combination of the two: 
\begin{shaded}
\begin{multline}\label{summaryAdSpmbcs}
     \Pi^{\text{AdS}}_{\alpha \Delta_+ + \beta \Delta_-,J}\left(z_1,{\bf k},u;z_2,-{\bf k},{\bar u}\right) = \underbrace{\csc\left(\pi\left(u+{\bar u}\right)\right)}_{\text{contact terms}}\underbrace{\left[ \alpha\,\omega_{\Delta_+}\left(u,{\bar u}\right)+\beta\,\omega_{\Delta_-}\left(u,{\bar u}\right)\right]}_{\text{boundary conditions}}\\ \times \Gamma(i\mu)\Gamma(-i\mu)\underbrace{\Omega^{\text{AdS}}_{\mu,J}\left(z_1,{\bf k},u;z_2,-{\bf k},{\bar u}\right)}_{\text{harmonic function}}.
\end{multline}
\end{shaded}

The Mellin-Barnes representation of in-in propagators in dS takes the same form. In particular, the most general dS bulk-bulk propagator for the $\pm {\hat \pm}$ branches of the in-in contour reads
\begin{shaded}
\begin{multline}\label{summarydSpm}
\hspace*{-0.5cm}    \Pi^{\text{dS}, \pm {\hat \pm}}_{\alpha \Delta_+ + \beta \Delta_-,J}\left(\eta_1,{\bf k},u;\eta_2,-{\bf k},{\bar u}\right) = \underbrace{\csc\left(\pi\left(u+{\bar u}\right)\right)}_{\text{contact terms}}\underbrace{\left[ \alpha_{\pm {\hat \pm}}\,\omega_{\Delta_+}\left(u,{\bar u}\right)+\beta_{\pm {\hat \pm}}\,\omega_{\Delta_-}\left(u,{\bar u}\right)\right]}_{\text{boundary conditions}}\\ \times \Gamma(i\mu)\Gamma(-i\mu)\underbrace{\Omega^{\text{dS}, \pm {\hat \pm}}_{\mu,J}\left(\eta_1,{\bf k},u;\eta_2,-{\bf k},{\bar u}\right)}_{\text{harmonic function}}, 
\end{multline}
\end{shaded}
\noindent where the $\pm {\hat \pm}$ harmonic function is the following analytic continuation of the AdS harmonic function \eqref{AdSharm}:
\begin{subequations}
\begin{align}
    \Omega^{\text{dS}, \pm, {\hat \pm}}_{\mu,J}\left(\eta_1,{\bf k};\eta_2,-{\bf k}\right) &=  \Omega^{\text{AdS}}_{\mu,J}\left(-\eta_1 e^{\pm \frac{i\pi}{2}},{\bf k};-\eta_2 e^{{\hat \pm} \frac{i\pi}{2}},-{\bf k}\right)\\
    &= \frac{\mu^2}{\pi} K^{\pm}_{\Delta_+,J}\left(\eta_1, {\bf k}\right)K^{{\hat \pm}}_{\Delta_-,J}\left(\eta_2, -{\bf k}\right),
\end{align}
\end{subequations}
where in the second equality we used the analytic continuations \eqref{analbubo} of the AdS bulk-boundary propagators.

The precise linear combination of $\Delta_\pm$ modes appearing in the dS in-in propagators \eqref{summarydSpm} is fixed by the choice of vacuum at early times, which determines the boundary (or late time) behaviour of the exchanged field. The Bunch Davies (B.D.) vacuum corresponds to the choice 
\begin{subequations}\label{BDab}
\begin{align}
    \alpha^{\text{B.D.}}_{+ +}&=+\frac{1}{8} \left[\cot \left(\pi \mu i\right)-i\right]\,,& \alpha^{\text{B.D.}}_{--}&=+\frac{1}{8} \left[\cot \left(\pi \mu i\right)+i\right]\,,\\
    \beta^{\text{B.D.}}_{+ +}&=-\frac{1}{8} \left[\cot \left(\pi \mu i\right)+i\right]\,,& \beta^{\text{B.D.}}_{--}&=-\frac{1}{8} \left[\cot \left(\pi \mu i\right)-i\right]\,,\\
    \alpha^{\text{B.D.}}_{\pm\mp}&=-\frac12\,\csc\left(\pi \mu i\right)\,,& \beta^{\text{B.D.}}_{\pm\mp}&=+\frac12\,\csc\left(\pi \mu i\right)\,.
\end{align}
\end{subequations}
The fact that these coefficients are different for each of the in-in propagators \eqref{summarydSpm} tells us that they cannot be regarded as analytic continuations of one and the same AdS bulk-bulk propagator \eqref{summaryAdSpmbcs}. Instead, the in-in propagators are analytic continuations of \emph{different} AdS bulk-bulk propagators i.e. with different $\alpha$s and $\beta$s. Similar statements about vacuum expectation values in dS not being analytic continuations of those in AdS have been made in previous works e.g. \cite{Bousso:2001mw,Spradlin:2001nb,Balasubramanian:2002zh}. We hope this work provides further clarification.

\section{Three-point (A)dS Contact Diagrams}
\label{section::3pt}

The Mellin-Barnes representation of a three-point contact diagram in (A)dS for fields of spin $J_1$-$J_2$-$J_3$  is defined as
\begin{equation}\label{3ptcormom}
   {\cal A}^{\text{(A)dS}}_{\mu_1,J_1;\mu_2,J_2;\mu_3,J_3}\left({\bf k}_1,{\bf k}_2,{\bf k}_3\right) = \int^{+i\infty}_{-i\infty} \left[ds\right]_3\,  {\cal A}^{\text{(A)dS}}_{\mu_1,J_1;\mu_2,J_2;\mu_3,J_3}\left({\bf k}_1,s_1;{\bf k}_2,s_2;{\bf k}_3,s_3\right),
\end{equation}
where $\left[ds\right]_3=\prod\limits^3_j\frac{ds_j}{2\pi i}$ and
\begin{multline}\label{MBrep3pt}
 \hspace*{-1cm} {\cal A}^{\text{(A)dS}}_{\mu_1,J_1;\mu_2,J_2;\mu_3,J_3}\left({\bf k}_1,s_1;{\bf k}_2,s_2;{\bf k}_3,s_3\right)  = \delta\left(\tfrac{x}{4}-s_1-s_2-s_3\right) \mathfrak{C}^{\text{(A)dS}}_{\mu_1,J_1;\mu_2,J_2;\mu_3,J_3}\left(s_1,s_2,s_3|\xi_k \cdot {\bf k}_j,\xi_k \cdot \xi_j\right)  \\  \times \rho_{\mu_1,\mu_2,\mu_3}\left(s_1,s_2,s_3\right)\prod^3_{j=1}\left(\frac{k_j}{2}\right)^{-2s_j+i\mu_j}.
\end{multline}
The function $\rho_{\mu_1,\mu_2,\mu_3}\left(s_1,s_2,s_3\right)$ carries two infinite families of poles for each Mellin variable,
\begin{equation}\label{rho3}
    \rho_{\mu_1,\mu_2,\mu_3}\left(s_1,s_2,s_3\right)=\prod^3_{j=1}\frac{1}{2\sqrt{\pi}}\Gamma\left(s_j+\tfrac{i\mu_j}{2}\right)\Gamma\left(s_j-\tfrac{i\mu_j}{2}\right).
\end{equation}
which are the poles in the Mellin-Barnes representation of the corresponding bulk-boundary propagators \eqref{MBbubosc}, which are Bessel functions \cite{Raju:2010by}. The function $\mathfrak{C}^{\text{(A)dS}}_{\mu_1,J_1;\mu_2,J_2;\mu_3,J_3}$ encodes the tensorial structure, which is a polynomial in the contractions $\left(\xi_k \cdot {\bf k}_j\right)$ and $\left(\xi_k \cdot \xi_j\right)$ with the boundary polarization vectors $\xi_i$, and is a rational function of the Mellin variables $s_j$.  In \cite{Sleight:2019hfp} we gave its explicit form for 3pt functions involving two scalars and a spin-$J$ field.

The Dirac delta function in the Mellin-Barnes representation enforces a constraint among the Mellin variables $s_i$ coming from the Dilatation Ward identity (see e.g. \cite{Bzowski:2013sza} for its form in momentum space), which requires
\begin{equation}
    s_1+s_2+s_3=\frac{x}{4}, \qquad x=d+N,\label{scalecond}
\end{equation}
where $N$ is the degree of the polynomial $\mathfrak{C}^{\text{(A)dS}}_{\mu_1,J_1;\mu_2,J_2;\mu_3,J_3}$ in the contractions $\left(\xi_k \cdot {\bf k}_j\right)$. For  three-point functions involving two scalars and a spin-$J$ field, for which the 3pt conformal structure is unique, this is $x=d+2J$ \cite{Sleight:2019hfp}. From a bulk perspective, the Dirac delta function can be expressed as an integral over the bulk radial co-ordinate, which for Poincar\'e co-ordinates \eqref{poincare} in EAdS reads:
\begin{equation}\label{ddeta}
  i \pi  \delta\left(\tfrac{x}{4}-s_1-s_2-s_3\right)= \lim_{z_0 \to 0}\int^{\infty}_{z_0}\frac{dz}{z}\, z^{\frac{x}{2}-2\left(s_1+s_2+s_3\right)}.
\end{equation}

Note that three-point contact diagrams in AdS and dS which are generated by the same cubic vertex are given by the same tensorial structure, differing only by the 3pt function coefficient. The precise proportionality constant between them can be worked out straightforwardly using the relation \eqref{dSAdSbubo} between the Mellin-Barnes representations of bulk-boundary propagators in AdS and dS, which differ by a simple phase. In particular, the $\pm$ branch three-point function in dS is obtained from its AdS counterpart by multiplying each leg by the phase converting it to a $\pm$ branch bulk-boundary propagator in dS. The resulting overall phase is a constant, as required by the Dilatation Ward identity \eqref{scalecond}:\footnote{The additional $\pm i$, which comes from analytic continuation of the bulk radial integral. For each external leg one should also multiply by the factor
\begin{equation}\label{Ncal}
  {\cal N}^{\pm}_{\frac{d}{2}+i\mu_j,J}\left(\eta_0\right)=\left(-\eta_0e^{\mp \frac{i\pi}{2}}\right)^{\tfrac{d}{2}+i\mu_j-J}\frac{\Gamma\left(-i\mu_j\right)}{2\sqrt{\pi}},
\end{equation}
coming from the boundary limit of the (Bunch Davies) bulk-bulk propagator \eqref{summarydSpm}. See \cite{Sleight:2019hfp} for the details.}
\begin{subequations}\label{pmdS3pt}
\begin{align}
{\cal A}^{\text{dS},\pm}_{\mu_i,J_i}\left({\bf k}_i,s_i\right) &=\pm i\,e^{\left[\mp \sum\limits^3_{j=1} \left(s_j+\frac{i\mu_j}{2}\right)\pi i\right]} {\cal A}^{\text{AdS}}_{\mu_i,J_i}\left({\bf k}_i,s_i\right)\\
&=\pm i\,e^{\left[\mp \left(\tfrac{x}{4}+\tfrac{i\left(\mu_1+\mu_2+\mu_3\right)}{2}\right) \pi i\right]} {\cal A}^{\text{AdS}}_{\mu_i,J_i}\left({\bf k}_i,s_i\right),
\end{align}
\end{subequations}
where in the second equality we used the constraint \eqref{scalecond}. The full in-in 3pt function is the sum of the two, giving
\begin{subequations}
\begin{align}
{\cal A}^{\text{dS}}_{\mu_i,J_i}\left({\bf k}_i,s_i\right)&={\cal A}^{\text{dS},+}_{\mu_i,J_i}\left({\bf k}_i,s_i\right)+{\cal A}^{\text{dS},-}_{\mu_i,J_i}\left({\bf k}_i,s_i\right)\\
&= 2 \sin \left[\left(\tfrac{x}{4}+\tfrac{i\left(\mu_1+\mu_2+\mu_3\right)}{2}\right) \pi\right] {\cal A}^{\text{AdS}}_{\mu_i,J_i}\left({\bf k}_i,s_i\right).\label{dSsinAdS}
\end{align}
\end{subequations}
This extends the results of \cite{Sleight:2019mgd,Sleight:2019hfp} to a general triplet of particles and cubic couplings. Previous works on connecting 3pt AdS contact diagrams and their in-in counterparts in dS include \cite{Maldacena:2002vr,McFadden:2011kk,Maldacena:2011nz,Bzowski:2011ab}.

\section{Four-point (A)dS Exchanges}

The on-shell exchange of a particle is represented in four-point functions on the boundary of (A)dS space by Casimir invariants of the Conformal group. In particular, the source-free equation of motion \eqref{harmadseom} satisfied by the corresponding on-shell propagator is equivalent to the quadratic Casimir equation 
\begin{equation}\label{casimirquad}
\Bigg\{\hat{\mathcal{C}}_2-\Big[J(d+J-2)-\Delta(d-\Delta)\Big]\Bigg\}\,f_{\Delta,J}\left({\bf k}_1,{\bf k}_2; {\bf k}_3,{\bf k}_4 \right)=0\,,
\end{equation}
where $\hat{\mathcal{C}}_2$ is the quadratic Casimir of the Conformal group. This is equivalent to the homogeneous conformal invariance condition for exchanges introduced in \cite{Arkani-Hamed:2015bza} in the context of the Cosmological Bootstrap. An on-shell exchange in (A)dS with general boundary conditions is given by the general solution to this equation with a normalisation that is fixed by on-shell factorisation. 

Since boundary correlators dual to scattering processes in AdS are single valued functions\footnote{Below we will show that this is true also for exchanges in dS.} of the cross ratios in the Euclidean regime it is natural to consider single-valued solutions ${\cal F}_{\mu,J}$ to the quadratic Casimir equations, which we refer to as Conformal Partial Waves \cite{Dobrev:1977qv}. These are the boundary counter parts of the bulk harmonic function $\Omega_{\mu,J}$.\footnote{Sometimes Conformal Partial Waves are themselves also referred to as harmonic functions, e.g. in \cite{Caron-Huot:2017vep,Simmons-Duffin:2017nub}. Single-valuedness of the CPW is equivalent to the absence of a short distance singularity in $\Omega
^{\text{AdS}}_{\mu,J}$.} This duality is made manifest by the following representation of Conformal Partial Waves, which is a specific product of three-point functions:
\begin{equation}
{\cal F}_{\mu,J}\left({\bf k}_1,{\bf k}_2; {\bf k}_3,{\bf k}_4 \right) = {\cal N} \langle {\cal O}_{\Delta_1}\left({\bf k}_1\right)  {\cal O}_{\Delta_2}\left({\bf k}_2\right) {\cal O}_{\Delta_+,J}\left({\bf k}\right) \rangle \langle {\cal O}_{\Delta_-,J}\left(-{\bf k}\right) {\cal O}_{\Delta_3}\left({\bf k}_3\right)  {\cal O}_{\Delta_4}\left({\bf k}_4\right)  \rangle, 
\end{equation}
with normalisation ${\cal N}$. For four-point exchanges in AdS we choose the normalisation\footnote{and, in the case of spinning external legs, a basis of three-point conformal structures.} that is consistent with on-shell factorisation, defining
\begin{equation}\label{CPWads}
{\cal F}^{\text{AdS}}_{\mu,J}\left({\bf k}_1,{\bf k}_2; {\bf k}_3,{\bf k}_4 \right) = {\cal A}^{\text{AdS}}_{\mu_1,\mu_2;\mu,J}\left({\bf k}_1,{\bf k}_2; {\bf k}\right) {\cal A}^{\text{AdS}}_{-\mu,J;\mu_3,\mu_4}\left(-{\bf k};{\bf k}_3,{\bf k}_4\right),
\end{equation}
which is the product of the three-point functions generated by the cubic vertices that mediate the exchange. From a bulk perspective this is obtained by taking the expression for the exchange in terms of propagators and replacing the bulk-bulk propagator with the split representation \eqref{AdSharm} of the harmonic function $\Omega_{\mu,J}$. To obtain the general solution to the conformal invariance equation \eqref{casimirquad} we employ the duality between ${\cal F}_{\mu,J}$ and $\Omega_{\mu,J}$, where at the level of the Mellin-Barnes representation the most general boundary condition is implemented by multiplying with an arbitrary linear combination of projectors \eqref{w+-}:
\begin{shaded}
\begin{multline}\label{onshellexchAdS}
 \text{Disc}_{|\bold{k}|^2}\left[{\cal A}^{\text{AdS}}_{\alpha \Delta_+ + \beta \Delta_-,J}\left({\bf k}_1,s_1,{\bf k}_2,s_2;{\bf k}_3,s_3,{\bf k}_4,s_4\right) \right] = \underbrace{\left[ \alpha\,\omega_{\Delta_+}\left(u,{\bar u}\right)+\beta\,\omega_{\Delta_-}\left(u,{\bar u}\right)\right]}_{\text{boundary conditions}}\\ \times \Gamma(i\mu)\Gamma(-i\mu)\underbrace{{\cal F}^{\text{AdS}}_{\mu,J}\left({\bf k}_1,s_1,{\bf k}_2,s_2;{\bf k}_3,s_3,{\bf k}_4,s_4\right)}_{\text{Conformal Partial Wave}}, 
\end{multline}
\end{shaded}
\noindent where the Mellin-Barnes representation of the Conformal Partial Wave is simply
\begin{equation}
\hspace*{-0.25cm}{\cal F}^{\text{AdS}}_{\mu,J}\left({\bf k}_{1,2},s_{1,2};{\bf k}_{3,4},s_{3,4}\right) = {\cal A}^{\text{AdS}}_{\mu_1,\mu_2;\mu,J}\left({\bf k}_{1,2},s_{1,2}; {\bf k},u\right) {\cal A}^{\text{AdS}}_{-\mu,J;\mu_3,\mu_4}\left(-{\bf k},{\bar u};{\bf k}_{3,4},s_{3,4}\right),
\end{equation}
which is inherited from the Mellin-Barnes representation \eqref{3ptcormom} of the three-point functions.

Above we fixed the form of a general on-shell exchange in AdS by solving the Casimir equation \eqref{casimirquad} and fixing the normalisation from on-shell factorisation. The full, off-shell, exchange satisfies the same equation but with a contact term source on the r.h.s.. This is equivalent to the equation of motion for the corresponding bulk-bulk propagator, which has a delta function source on the r.h.s.. To obtain the full exchange one possibility is to directly solve the conformal invariance equation \eqref{casimirquad} with a source \cite{Arkani-Hamed:2015bza}, as was done in \cite{Arkani-Hamed:2018kmz} in the context of the Cosmological Bootstrap. Another possibility is to try to reconstruct the solution to this equation from the knowledge of the solution to its homogeneous counterpart \eqref{casimirquad} i.e. to reconstruct the full exchange from its on-shell part \eqref{onshellexchAdS}. This is closer in spirit to generalised unitarity methods both in amplitudes \cite{Bern:2011qt} and the conformal bootstrap \cite{Caron-Huot:2017vep,Alday:2017gde,Meltzer:2019nbs}. As we saw in \S \ref{sec::prop} this is straightforward at the level of the Mellin-Barnes representation, where the bulk-bulk propagator is obtained from its on-shell part by multiplying it with $\csc\left(\pi\left(u+{\bar u}\right)\right)$. The full exchange is therefore:
\begin{shaded}
\begin{multline}
{\cal A}^{\text{AdS}}_{\alpha \Delta_+ + \beta \Delta_-,J}\left({\bf k}_{1,2},s_{1,2};{\bf k}_{3,4},s_{3,4}\right) = \underbrace{\csc\left(\pi\left(u+{\bar u}\right)\right)}_{\text{contact terms}}\underbrace{\left[ \alpha\,\omega_{\Delta_+}\left(u,{\bar u}\right)+\beta\,\omega_{\Delta_-}\left(u,{\bar u}\right)\right]}_{\text{boundary conditions}}\\ \times\Gamma(i\mu)\Gamma(-i\mu) \underbrace{{\cal F}^{\text{AdS}}_{\mu,J}\left({\bf k}_{1,2},s_{1,2};{\bf k}_{3,4},s_{3,4}\right)}_{\text{Conformal Partial Wave}} .
\end{multline}
\end{shaded}

For exchanges in dS we proceed in exactly the same way: We first solve the homogeneous conformal invariance equation \eqref{casimirquad} for the on-shell $\pm {\hat \pm}$ exchange, from which we can straightforwardly reconstruct the full exchange. For the $\pm {\hat \pm}$ exchange the conformal partial wave is normalised as
\begin{equation}\label{CPWds}
{\cal F}^{\text{dS}, \pm, {\hat \pm}}_{\mu,J}\left({\bf k}_1,{\bf k}_2; {\bf k}_3,{\bf k}_4 \right) = {\cal A}^{\text{dS}, \pm}_{\mu_1,\mu_2;\mu,J}\left({\bf k}_1,{\bf k}_2; {\bf k}\right) {\cal A}^{\text{dS}, {\hat \pm}}_{-\mu,J;\mu_3,\mu_4}\left(-{\bf k};{\bf k}_3,{\bf k}_4\right),
\end{equation}
which is a product of the $\pm$ and ${\hat \pm}$ three-point functions in dS. Following the above we can then immediately write down the general solution for the full dS exchange:
\begin{shaded}
\begin{multline}
\hspace*{-0.5cm}{\cal A}^{\text{dS}}_{\alpha \Delta_+ + \beta \Delta_-,J}\left({\bf k}_{1,2},s_{1,2};{\bf k}_{3,4},s_{3,4}\right) = \sum_{\pm {\hat \pm}}\underbrace{\csc\left(\pi\left(u+{\bar u}\right)\right)}_{\text{contact terms}}\underbrace{\left[ \alpha_{\pm {\hat \pm}}\,\omega_{\Delta_+}\left(u,{\bar u}\right)+\beta_{\pm {\hat \pm}}\,\omega_{\Delta_-}\left(u,{\bar u}\right)\right]}_{\text{boundary conditions}}\\ \times \Gamma(i\mu)\Gamma(-i\mu)\underbrace{{\cal F}^{\text{dS}, \pm {\hat \pm}}_{\mu,J}\left({\bf k}_{1,2},s_{1,2};{\bf k}_{3,4},s_{3,4}\right)}_{\text{Conformal Partial Wave}} ,
\end{multline}
\end{shaded}
\noindent where the full in-in exchange is the sum of $\pm {\hat \pm}$ exchanges. Recall that the coefficients  $\alpha_{\pm {\hat \pm}}$ and $\beta_{\pm {\hat \pm}}$ parameterise the choice of vacuum at early times, which for the Bunch Davies vacuum corresponds to the choice \eqref{BDab}.

Let us emphasise that the CPW for dS exchanges \eqref{CPWds} is proportional to that \eqref{CPWads} for exchanges in AdS since it amounts to a choice of normalisation that is determined by the requirement of on-shell factorisation. The precise proportionality constant is given by the proportionality constants between the $\pm$ and ${\hat \pm}$ three-point functions in dS and their AdS counterparts, which were given in \S \ref{section::3pt}. In particular,
\begin{equation}
{\cal F}^{\text{dS}, \pm, {\hat \pm}}_{\mu,J}\left({\bf k}_1,{\bf k}_2; {\bf k}_3,{\bf k}_4 \right) = e^{(d+2 J+i (\mp\mu_1\mp\mu_2\hat{\mp}\mu_3\hat{\mp}\mu_4))\frac{\pi i}{2}}{\cal F}^{\text{AdS}}_{\mu,J}\left({\bf k}_1,{\bf k}_2; {\bf k}_3,{\bf k}_4 \right).
\end{equation}
This means that we can express any in-in exchange in dS as a sum of AdS exchanges ${\cal A}^{\text{AdS}}_{\Delta_\pm,J}$. For the dS exchange in the Bunch Davies vacuum we have:\footnote{Note that, as in section \ref{section::3pt}, for each external leg one should multiply by the factor \eqref{Ncal}.}
\begin{shaded}
\begin{align}\label{dSexch}
    {\cal A}^{\text{dS},\, \text{B.D.}}_{\mu,J}&=\frac{1}{2}\, \csc \Big(\tfrac{\pi}{2}(2 \Delta_--d )\Big)\sin \left(\tfrac{\pi}{2} (\Delta_{1+2}+\Delta_-+J-d)\right) \sin \left(\tfrac{\pi}{2}(\Delta_{3+4}+\Delta_-+J-d)\right) {\cal A}^{\text{AdS}}_{\Delta_-,J}\nonumber\\
    &+\frac{1}{2}\, \csc \Big(\tfrac{\pi}{2}(2 \Delta_+-d )\Big)\sin \left(\tfrac{\pi}{2}  (\Delta_{1+2}+\Delta_++J-d)\right) \sin \left(\tfrac{\pi}{2}(\Delta_{3+4}+\Delta_++J-d)\right){\cal A}^{\text{AdS}}_{\Delta_+,J}\,,
\end{align}
\end{shaded}
\noindent where $\Delta_{i+j}\equiv\Delta_i+\Delta_j$.
This equation is the main result of this work. Let us emphasise that, although this relation between dS and AdS exchanges was derived by exploiting the Mellin-Barnes representation for momentum space, it holds also in position space. In this way we can directly import to dS space the techniques and results enjoyed by both the position and momentum space exchanges in AdS, some of which we shall explore in the following section. 

Note that the above relation between dS and AdS exchanges reproduces the OPE limit of the dS exchange derived in \cite{Sleight:2019mgd,Sleight:2019hfp}. The factors multiplying each AdS exchange are the products of sinusoidal factors \eqref{dSsinAdS} from the corresponding 3pt contact diagrams. A posteriori, one could have recovered the full expression for the dS exchange from its OPE limit by completing each single-trace conformal block to its corresponding AdS exchange. Assuming single-valuedness of the dS exchange, the result \eqref{dSexch} would have then been the unique possible solution. The way we arrived to the expression \eqref{dSexch} indeed confirms that single valuedness of dS exchanges would have been a valid assumption.

\section{Spectral representation, Mellin amplitudes and Crossing in de Sitter}

In the following we shall use the result \eqref{dSexch} to establish the following: In \S \ref{subsec::spectralrep} the position space spectral representation for dS exchanges, which for AdS exchanges was given in \cite{Cornalba:2007fs,Costa:2014kfa}. In \S \ref{subsec::Mellinamplitudes} we will then define the Mellin amplitudes \cite{Mack:2009mi} for tree-level four-point functions in dS from their AdS counterparts introduced in \cite{Penedones:2010ue,Paulos:2011ie,Fitzpatrick:2011ia}. In \S \ref{subsec::crossing} we shall write down the crossing decomposition of dS exchanges in terms of known results for the conformal block and crossing decomposition of their AdS counterparts. We briefly comment on the Polyakov-Bootstrap \cite{Polyakov:1974gs,Gopakumar:2016cpb} for dS.

\subsection{Spectral representation}
\label{subsec::spectralrep}

The spectral representation decomposition of a conformal 4pt function reads \cite{Costa:2012cb,Caron-Huot:2017vep,Simmons-Duffin:2017nub}:
\begin{align}
    \left\langle\mathcal{O}_1\mathcal{O}_2\mathcal{O}_3\mathcal{O}_4\right\rangle=(\text{non-norm.})+\sum_{J}\int_{-\infty}^\infty d\nu\, a(\nu)\,{\cal F}_{\nu,J}\,,
\end{align}
where the spectral integral captures only the normalisable part of the 4pt function and $a(\nu)$ has poles at the scaling dimensions of physical exchanged primary operators. The spectral representation of spinning AdS exchanges was given in \cite{Costa:2014kfa} (and later on in \cite{Sleight:2017fpc} for spinning external legs), which follows from the harmonic function decomposition \eqref{specpropAdSdirich} and \eqref{specpropAdSdeu} of the bulk-bulk propagators. From the relation \eqref{dSexch} we can then immediately write down the spectral representation for exchanges in dS:
\begin{shaded}
\begin{multline}\label{spectraldS}
    {\cal A}^{\text{dS},\, \text{B.D.}}_{\mu,J}=-\,\underbrace{\sin \left(\tfrac{\pi}{2} (\Delta_1+\Delta_2+\Delta_-+J-d)\right) \sin \left(\tfrac{\pi}{2} (\Delta_3+\Delta_4+\Delta_-+J-d)\right)\,{\cal F}^{\text{AdS}}_{\mu,J}}_{\text{non-normalisable}}\\\frac{1}2\sin\left(\tfrac{\pi}{2} (\Delta_1+\Delta_2+\Delta_3+\Delta_4+2J-d)\right)\int_{-\infty}^\infty d\nu\,\frac1{\nu^2+\left(\Delta_+-\frac{d}{2}\right)^2}\,{\cal F}^{\text{AdS}}_{\nu,J}\,,
\end{multline}
\end{shaded}
\noindent where the above spectral representation also formally extends to $\Delta_+$ on the principal series using the appropriate $\epsilon$-prescription, as was shown in \S \tcb{4.7} of \cite{Sleight:2019hfp}.

Since it is a solution to the homoegeneous conformal invariance equation \eqref{casimirquad}, the CPW on the first line of \eqref{spectraldS} does not contain bulk contact terms. These are encoded in the integral over the spectral parameter $\nu$, from the residues of poles in the AdS 3pt function coefficients. The relation \eqref{spectraldS} (and the expression \eqref{dS_mellin} for the corresponding Mellin amplitude) tell us that the 4pt contact terms are given by the AdS counterparts multiplied by the factor $\sin\left(\tfrac{\pi}{2} (2J-d+\sum\limits^4_{j=1}\Delta_j)\right)$. This was also observed in \cite{Sleight:2019mgd,Sleight:2019hfp} and extends to 4pts the relation \eqref{dSsinAdS} between 3pt AdS and dS contact diagrams.

\subsection{Mellin amplitudes}
\label{subsec::Mellinamplitudes}

The Mellin amplitude for the normalisable AdS exchange ${\cal A}^{\text{AdS}}_{\Delta_+,J}$ is given by \cite{Penedones:2010ue,Costa:2012cb}:\footnote{See also \cite{Chen:2017xdz,Sleight:2018epi} on spinning external legs.}
\begin{align}
   {\cal M}^{\text{AdS}}_{\Delta_+,J}\left(s,t\right) = C_{\Delta_1,\Delta_2,\Delta_+;J}C_{\Delta_3,\Delta_4,\Delta_+;J} \sum_{m=0}^{\infty}\frac{\mathcal{Q}_{\tau_+,J|m}(s)}{t-\tau_+-2m}+\text{contact}\,,
\end{align}
where $\tau_+ = \Delta_+-J$ is the twist and $C_{\Delta_i,\Delta_j,\Delta_+;J}$ are the OPE coefficients.\footnote{\label{adsope}More precisely these are $C_{\Delta_i,\Delta_j,\Delta_+;J}=\mathsf{B}_{\Delta_i,\Delta_j,\Delta_+;J}/\sqrt{C_{\Delta_+,J}}$. The factors $\mathsf{B}_{\Delta_i,\Delta_j,\Delta_\pm;J}$ arise from the bulk integration (see equation (131) of \cite{Costa:2014kfa}) and $C_{\Delta,J}$ is the 2pt function coefficient.} The $\mathcal{Q}_{\tau_+,J|m}(s)$ are kinematic polynomials. We follow the conventions and notations of \cite{Costa:2012cb}, to which we refer the reader for details. The contact terms are simply polynomials in the Mellin variables $s, t$, of degree no greater than $J-1$. The Mellin amplitude associated to the Conformal Partial Wave \eqref{CPWads} is instead given by \cite{Dolan:2011dv}
\begin{subequations}
\begin{align}
    {\cal F}^{\text{AdS}}_{\mu,J}\left(s,t\right)&=\mathcal{N}_{\tau_+,J}\,\mathsf{B}_{\Delta_1,\Delta_2,\Delta_+;J}\mathsf{B}_{\Delta_3,\Delta_4,\Delta_-;J}\Omega_{J}(t) P_{\tau_+,J}(s,t)\,,\\
    \Omega_{J}(t)&=\frac{\Gamma\left(\frac{\tau_+-t}2\right)\Gamma\left(\frac{d-2J-t-\tau_+}2\right)}{\Gamma\left(\tfrac{-t+\Delta_1+\Delta_2}2\right)\Gamma\left(\tfrac{-t+\Delta_3+\Delta_4}2\right)}\,,\\
    \mathcal{N}_{\tau_+,J}&=\frac{1}{\Gamma \left(\tfrac{\tau_+ +\Delta_1-\Delta_2}{2}\right) \Gamma \left(\tfrac{\tau_+ -\Delta_1+\Delta_2}{2}\right) \Gamma \left(\tfrac{d-\tau_+ +\Delta_3-\Delta_4}{2}\right) \Gamma \left(\tfrac{d-\tau_+ -\Delta_3+\Delta_4}{2} \right)}
\end{align}
\end{subequations}
where $P_{\tau_+,J}(s,t)$ are the so-called Mack polynomials \cite{Mack:2009mi,Costa:2012cb}. For the latter we use the normalisation employed in \cite{Sleight:2018epi}.

From \eqref{spectraldS} we can then immediately write down the  Mellin amplitude for the dS exchange \eqref{dSexch}:
\begin{shaded}
\begin{multline}\label{dS_mellin}
    {\cal M}^{\text{dS},\, \text{B.D.}}_{\mu,J}\left(s,t\right)=\frac{1}2\sin\left(\tfrac{\pi}{2} (\Delta_1+\Delta_2+\Delta_3+\Delta_4+2J-d)\right){\cal M}^{\text{AdS}}_{\Delta_+,J}\left(s,t\right)\\-\,\sin \left(\tfrac{\pi}{2} (\Delta_1+\Delta_2+\Delta_-+J-d)\right) \sin \left(\tfrac{\pi}{2} (\Delta_3+\Delta_4+\Delta_-+J-d)\right)\,{\cal F}^{\text{AdS}}_{\mu,J}\left(s,t\right)\,.
\end{multline}
\end{shaded}

Note that this result unlocks another tool for studying the flat limit of dS correlators. So far, their flat limit has been explored mostly in momentum space by taking ${k_t \equiv |{\bf k}_1|+|{\bf k}_2|+|{\bf k}_3|+|{\bf k}_4| \to 0}$, which recovers the high energy limit of the corresponding flat space amplitude \cite{Maldacena:2011nz,Raju:2012zr}. With the result \eqref{dS_mellin}, one can now also try to make use of Penedones' formula \cite{Penedones:2010ue} relating a Mellin amplitude in a {large-$N$} CFT and the corresponding flat space S-Matrix. For dS, naively one would write:
\begin{equation}
\lim_{R\to\infty}\int
_{-i\infty}^{+i\infty}d\alpha\ e^\alpha \alpha^{\frac{d-\Delta_1-\Delta_2-\Delta_3-\Delta_4}{2}}\,{\cal M}^{\text{dS},\, \text{B.D.}}_{\mu,J}\left(\tfrac{R^2s}{4\alpha},\tfrac{R^2t}{4\alpha};\Delta_j=iR m_j\right)\,,
\end{equation}
up to an overall normalisation following from that of the bulk-boundary propagators.

\subsection{Conformal Block Decomposition and Crossing}
\label{subsec::crossing}

Using \eqref{dSexch} the conformal block decomposition of dS exchanges can be written down from the knowledge of the decomposition of their AdS counterparts ${\cal A}^{\text{AdS}}_{\Delta_\pm,J}$. The decomposition of the ${\sf s}$-channel dS exchange \eqref{dSexch} in terms of direct channel conformal blocks reads:
\begin{align}\label{directCB}
    {}^{({\sf s})}{\cal A}^{\text{dS},\, \text{B.D.}}_{\mu,J}(u,v)&=a_{\Delta_+,J}g_{\Delta_+,J}(u,v)+a_{\Delta_-,J}g_{\Delta_-,J}(u,v)\\&+\sum_{n,J^\prime=0}^J\left[{}^{({\sf s})}a_{n,J^\prime|\Delta_+,J}+{}^{({\sf s})}a_{n,J^\prime|\Delta_-,J}\right]\, g_{\Delta_1+\Delta_2+2n,J^\prime}(u,v)\nonumber\\&+\sum_{n,J^\prime=0}^J \left[{}^{({\sf s})}b_{n,J^\prime|\Delta_+,J}+{}^{({\sf s})}b_{n,J^\prime|\Delta_-,J}\right]\, g_{\Delta_3+\Delta_4+2n,J^\prime}(u,v)\,,\nonumber
\end{align}
where the OPE coefficients are given in terms of their AdS counterparts via
\begin{multline}
     \frac{a_{\Delta_{\pm},J}}{a_{\Delta_{\pm},J}^{\text{AdS}}}=\frac{{}^{({\sf s})}a_{n,J^\prime|\Delta_\pm,J}}{{}^{({\sf s})}a^{\text{AdS}}_{n,J^\prime|\Delta_\pm,J}}=\frac{{}^{({\sf s})}b_{n,J^\prime|\Delta_\pm,J}}{{}^{({\sf s})}b^{\text{AdS}}_{n,J^\prime|\Delta_\pm,J}}\\=\frac{1}{2}\, \csc \Big(\tfrac{\pi}{2}(2 \Delta_\pm-d )\Big)\sin \left(\tfrac{\pi}{2}  (\Delta_{1+2}+\Delta_\pm+J-d)\right) \sin \left(\tfrac{\pi}{2}(\Delta_{3+4}+\Delta_\pm+J-d)\right).\nonumber 
\end{multline}
For the single-trace exchanges $a_{\Delta_\pm,J}$ on the first line of \eqref{directCB}, the AdS OPE coefficients are simply (\emph{cf.} footnote \ref{adsope})
\begin{equation}
    a_{\Delta_{\pm},J}^{\text{AdS}}=\frac{\mathsf{B}_{\Delta_1,\Delta_2,\Delta_\pm;J}\mathsf{B}_{\Delta_3,\Delta_4,\Delta_\pm;J}}{C_{\Delta_\pm,J}}.
\end{equation}
The AdS OPE coefficients $a^{\text{AdS}}_{n,J^\prime|\Delta_\pm,J}$ and $b^{\text{AdS}}_{n,J^\prime|\Delta_\pm,J}$ for exchanges of double-trace operators on the second and third lines of \eqref{directCB} are non-analytic in spin $J^\prime$. These have been computed e.g. in \cite{Gopakumar:2016cpb,Alday:2017gde,Zhou:2018sfz,Gopakumar:2018xqi,Sleight:2019ive}.

We can also determine the conformal block decomposition of exchanges under crossing. The decomposition of a ${\sf t}$- or ${\sf u}$-channel dS exchange in terms of direct channel conformal blocks reads
\begin{align}
     {}^{({\sf t, u})}{\cal A}^{\text{dS},\, \text{B.D.}}_{\mu,J}(u,v)&=\sum_{n,J^\prime=0}^\infty \left[{}^{({\sf t, u})}a^{\text{dS}}_{n,J^\prime|\Delta_+,J}+{}^{({\sf t, u})}a^{\text{dS}}_{n,J^\prime|\Delta_-,J}\right]\, g_{\Delta_1+\Delta_2+2n,J^\prime}(u,v)\\&+\sum_{n,J^\prime=0}^\infty \left[{}^{({\sf t, u})}b^{\text{dS}}_{n,J^\prime|\Delta_+,J}+{}^{({\sf t, u})}b^{\text{dS}}_{n,J^\prime|\Delta_-,J}\right]\, g_{\Delta_3+\Delta_4+2n,J^\prime}(u,v)\,, \nonumber
\end{align}
where now
\begin{multline}
     \frac{{}^{({\sf t})}a_{n,J^\prime|\Delta_\pm,J}}{{}^{({\sf t})}a^{\text{AdS}}_{n,J^\prime|\Delta_\pm,J}}=\frac{{}^{({\sf t})}b_{n,J^\prime|\Delta_\pm,J}}{{}^{({\sf t})}b^{\text{AdS}}_{n,J^\prime|\Delta_\pm,J}}\\=\frac{1}{2}\, \csc \Big(\tfrac{\pi}{2}(2 \Delta_\pm-d )\Big)\sin \left(\tfrac{\pi}{2} (\Delta_{1+3}+\Delta_\pm+J-d)\right) \sin \left(\tfrac{\pi}{2}(\Delta_{2+4}+\Delta_\pm+J-d)\right),\nonumber 
\end{multline}
and
\begin{multline}
     \frac{{}^{({\sf u})}a_{n,J^\prime|\Delta_\pm,J}}{{}^{({\sf u})}a^{\text{AdS}}_{n,J^\prime|\Delta_\pm,J}}=\frac{{}^{({\sf u})}b_{n,J^\prime|\Delta_\pm,J}}{{}^{({\sf u})}b^{\text{AdS}}_{n,J^\prime|\Delta_\pm,J}}\\=\frac{1}{2}\, \csc \Big(\tfrac{\pi}{2}(2 \Delta_\pm-d )\Big)\sin \left(\tfrac{\pi}{2} (\Delta_{1+4}+\Delta_\pm+J-d)\right) \sin \left(\tfrac{\pi}{2}(\Delta_{2+3}+\Delta_\pm+J-d)\right).\nonumber 
\end{multline}
The AdS OPE coefficients ${}^{({\sf t,  u})}a^{\text{AdS}}_{n,J^\prime|\Delta_\pm,J}$ and ${}^{({\sf t,  u})}b^{\text{AdS}}_{n,J^\prime|\Delta_\pm,J}$ of double-trace operators induced by crossed channel exchanges have a universal component that is analytic in spin $J^\prime$, which have been computed in various works \cite{Gopakumar:2016cpb,Alday:2017gde,Sleight:2018ryu,Liu:2018jhs,Zhou:2018sfz,Albayrak:2019gnz,Gopakumar:2018xqi}. In \cite{Sleight:2018ryu} they were in particular shown to be given by finite sums of Wilson functions. In addition to the analytic in spin part there are also non-analytic in spin contributions coming from bulk contact terms, which have been computed e.g. in \cite{Heemskerk:2009pn,Gopakumar:2016cpb,Alday:2017gde,Zhou:2018sfz,Gopakumar:2018xqi,Sleight:2019ive}. 

Regarding bulk exchanges as a basis to decompose CFT 4pt functions in the spirit of the Polyakov Bootstrap \cite{Polyakov:1974gs,Gopakumar:2016cpb}, the freedom to add bulk contact terms is a source of ambiguity in the OPE data. A prescription to fix this contact term ambiguity for CFT correlators dual to physics in AdS was proposed in \cite{Mazac:2019shk,Sleight:2019ive} (see also the related \cite{Penedones:2019tng}). Given our result \eqref{dSexch}, we expect this prescription and the corresponding (cyclic) Polyakov Blocks to carry over to correlators in Euclidean CFTs dual to physics in dS, although their relevance at the non-perturbative level in dS should be clarified.

\section*{Acknowledgments}

The research of C.S. is supported by the European Union's Horizon 2020 research and innovation programme under the Marie Sk\l odowska-Curie grant agreement No 793661. The research of M.T. was partially supported by the program  ``Rita  Levi  Montalcini'' of the MIUR (Minister for Instruction, University and Research) and the INFN initiative STEFI.

\bibliographystyle{JHEP}
\bibliography{refs}

\end{document}